# A Compact and Interpretable Convolutional Neural Network for Cross-Subject Driver Drowsiness Detection from Single-Channel EEG

Jian Cui
*Fraunhofer Singapore*
*Nanyang Technological University*
Singapore
cuijian@ntu.edu.sg

Zirui Lan, Yisi Liu
*Fraunhofer Singapore*
Singapore
lan.zirui@fraunhofer.sg,
liu.yisi@fraunhofer.sg

Ruilin Li
*Nanyang Technological University*
Singapore
RUILIN001@e.ntu.edu.sg

Fan Li
*Fraunhofer Singapore*
*Nanyang Technological University*
Singapore
lifan@ntu.edu.sg

Olga Sourina
*Fraunhofer Singapore*
*Nanyang Technological University*
Singapore
EOSourina@ntu.edu.sg

Wolfgang Müller-Wittig
*Fraunhofer Singapore*
*Nanyang Technological University*
Singapore
Wolfgang.Mueller-wittig@fraunhofer.sg

*Abstract*—Driver drowsiness is one of the main factors leading to road fatalities and hazards in the transportation industry. Electroencephalography (EEG) has been considered as one of the best physiological signals to detect drivers' drowsy states, since it directly measures neurophysiological activities in the brain. However, designing a calibration-free system for driver drowsiness detection with EEG is still a challenging task, as EEG suffers from serious mental and physical drifts across different subjects. In this paper, we propose a compact and interpretable Convolutional Neural Network (CNN) to discover shared EEG features across different subjects for driver drowsiness detection. We incorporate the Global Average Pooling (GAP) layer in the model structure, allowing the Class Activation Map (CAM) method to be used for localizing regions of the input signal that contribute most for classification. Results show that the proposed model can achieve an average accuracy of 73.22% on 11 subjects for 2-class cross-subject EEG signal classification, which is higher than conventional machine learning methods and other state-of-art deep learning methods. It is revealed by the visualization technique that the model has learned biologically explainable features, e.g., Alpha spindles and Theta burst, as evidence for the drowsy state. It is also interesting to see that the model uses artifacts that usually dominate the wakeful EEG, e.g., muscle artifacts and sensor drifts, to recognize the alert state. The proposed model illustrates a potential direction to use CNN models as a powerful tool to discover shared features related to different mental states across different subjects from EEG signals. ***Codes available from:*** **https://github.com/cuijiancorbin/A-Compact-and-Interpretable-Convolutional-Neural-Network-for-Single-Channel-EEG**

***Extracted dataset available from:*** **https://figshare.com/articles/dataset/EEG_driver_drowsiness_dataset/14273687**

I. INTRODUCTION

Up to now, driver drowsiness is still a major cause of road fatalities in the transportation industry. As revealed by National Highway Traffic Safety Administration (NHTSA), over 72,000 police-reported crashes from 2009 to 2013 and 775 fatal accidents (2.1 percent of total fatalities) in 2018 were caused by tired driving [1]. Development of a real-time drowsiness state monitoring system is desirable for transport accident prevention. Actually, drowsiness can be detected from changes of various physiological signals and Electroencephalography (EEG) is a one of the best physiological signals to detect drivers' drowsy states, since it directly measures neurophysiological activities in the brain. It has been found that the oscillation patterns of EEG signals are strongly correlated with drowsiness [2, 3]. In addition, band power related features [4], time domain features, including typical entropy features [5] and multi-scale entropy features [6], have also been found useful for drowsiness detection from EEG signals. However, it is still a challenging task to build a calibration-free drowsiness monitoring system with EEG, because there exists a large variability of EEG signals among different subjects and even in different sessions within the same subject due to undesired mental or physical drifts. The non-stationary and low signal-to-noise rate characteristics of EEG add difficulty to the recognition task.

Deep learning has received much research attention in recent years. Since its initial success in many challenging image classification problems [7, 8], it has been rapidly developed and achieved successes in many domains, such as speech recognition, sensor readings, motion capture, spectrographs, electrocardiogram, electric devices, and simulated data [9]. Deep learning has also become a hot topic in the field of EEG signal processing. Convolutional Neural Network (CNN), as one of the major deep learning models, is the most widely used structure for EEG signal classification, as surveyed by Roy et al. [10] on 156 papers with relation to deep learning in this field. CNN allows end-to-end learning of essential characteristics from multi-channel high-dimensional EEG data in an incremental manner without the need for a priori feature crafting, which makes it potentially a powerful tool to discover high-level drowsiness-related features from a diversity of EEG signals across different subjects.

In this paper, we use CNN to discover and visualize cross-subject EEG features that can indicate drowsiness from single-channel EEG signals. Different from most existing works that treat deep learning models as black-box classifiers, we want to build a model that not only has a high classification accuracy but also can explain how it makes the decision, so that we are able to derive insights on what characteristics of EEG signals have been learned by the model. In this connection, we propose a compact and interpretable CNN model for driver drowsiness detection from single-channel EEG signals. We take advantage of Global Average Pooling layer [11] in the structure, which not only can largely reduce the amount of model parameters but also allows important parts of the input data for classification to be revealed by the Class Activation Map (CAM) method [12]. The proposed model can be potentially used as a powerful tool to discover interesting neurophysiological patterns from single-channel EEG signals.

The following parts of this paper are organized as follows. Related works are surveyed in Section II. Data preparation is described in Section III. The model structure and visualization techniques are illustrated in section IV. The model is compared with other state-of-the-art methods in Section V. The results are presented in Section VI, which is followed by discussion and future works in Section VII. Conclusions are made in Section VIII.

## II. Related works

EEG has been considered as an affordable brain imaging technique for recognition of different mental states, such as different emotions [13-16], different levels of mental workload [17], stress [17, 18], and fatigue [19, 20]. It uses a set of electrodes to measure voltage fluctuations on the scalp, which are resulted from partial synchrony of local field potentials in many distinct cortical domains—each domain being, in the simplest case, a patch of cortex or unknown extent [21]. The macroscopic synchronized neuron firing patterns resulted from feedback connections between the neurons in one or multiple cortex patch areas give rise to oscillations of EEG signals in specific frequency bands. There are many studies showing the association between drowsiness and the oscillation patterns of EEG signals. For example, Torsvall [2] conducted an experiment involving 11 drivers and found that the EEG band power in the frequency range between 0.9 Hz and 11.9 Hz is higher during night driving than that during day driving. Lal et al. [3] found sleep deprivation can cause an increase in the EEG frequency band power between 0 Hz and 20 Hz. It was concluded by Klimesch et al. [22] that an increase in lower alpha power (6–10 Hz) can reflect the situation when subjects are struggling to sustain wakefulness while they are sleepy.

For driver state monitoring on daily basis, consumer-grade EEG devices with one or multiple electrodes are preferable over medical-grade ones due to their low price and ease of set-up. Conventional methods for drowsiness recognition rely on feature extraction from EEG signals, which requires expertise and/or a priori knowledge on the data in order to model some special EEG characteristics of interest. For example, Ogino et al. [23] compared three methods for feature extraction, which are Power Spectral Density (PSD), Autoregressive (AR) modeling, and Multiscale Entropy (MSE). They achieved a classification accuracy of 72.7% using the PSD features with Step-wise Linear Discriminant Analysis (SWLDA) for feature selection and Support Vector Machine (SVM) for classification. Lin et al. [24] used an unsupervised algorithm, which estimates level of drowsiness by measuring the deviation of the EEG spectra in the Theta and Alpha rhythm from that recorded in the alert state, for drowsiness detection from a single EEG channel placed at the occipital area. Silveira et al. [25] performed Discrete Wavelet Transform (DWT) on single-channel EEG signals and selected the most significant M-terms from the wavelet expansion as features for classification. Considering that the alpha frequency contains personal information about age and memory performance of a subject, Belakhdar et al. [26] proposed to use PSD calculated for individual alpha frequency (IAF) to model the extent of drowsiness. Venkat and Chinara [27] proposed a drowsiness detection model using time-domain features extracted by utilizing Wavelet Packet Transform (WPT).

By transforming the EEG signals into an array of feature vectors, conventional methods based on manual feature extraction may exclude some important characteristics of EEG signals that could be essential to drowsiness recognition. In comparison to traditional methods based on feature crafting, deep learning has advantage in automatically capturing essential characteristics from data—such models can directly learn a cascade of representations from raw and high-dimensional data by adjusting the parameters through back propagation. Deep learning methods are receiving more and more attention in the field of EEG signal processing, and they have been used for brain-computer interface (BCI) [28], recognition of different sleep stages [29], and classification of different workload levels [30]. For single-channel EEG signal classification, Akara et al. [31] proposed a deep learning model called DeepSleepNet for sleep stage classification. The model consists of a convolutional network part for feature extraction, and a bidirectional long short-term memory structure to learn transition rules among sleep stages. Sors et al. [32]

proposed a convolutional neural network containing 14 layers to deal with 30-s EEG epochs for sleep stage prediction. Yıldırım et al. [33] proposed a CNN model with 23 layers for automatic recognition of normal and abnormal single-channel EEG signals. Fahimi et al. [34] explored the use of an end-to-end deep CNN to detect the attentive mental state from single-channel raw EEG data. Bresch et al. [35] proposed a deep recurrent neural network for sleep stage classification. The model consists of cascaded convolutional layers followed by a max pooling layer and Long Short Term Memory (LSTM) layers in the structure, and achieved an agreement with a human annotator of Cohen's Kappa of ~0.73 using a training data set of 19 subjects. Mousavi et al. [36] considered to use a CNN model consisting of 9 convolutional layers followed by two fully connected layers for sleep stage classification. Ding et al. [37] employed cascaded CNN and attention mechanism to build a deep learning model for drivers drowsiness detection from single-channel EEG captured by mobile devices.

However, existing works mostly treat deep learning models as black-box classifiers, while their potential to be used for discovering and visualizing shared EEG features across different subjects is still underexplored. In fact, deriving insights on what characteristics of the EEG data that has been learned by the deep learning networks may be as important as achieving high classification accuracy, since it not only allows evaluation on the model but also potentially helps discover new neurophysiological patterns. Therefore, in this paper we propose to use deep learning as a powerful tool to discover interesting features that can explain drowsiness states from EEG signals of different subjects. For this purpose, we build a compact yet powerful CNN model with only essential components for the ease of interpretation.

III. DATA PREPARATION

*A. Dataset description*

An open EEG dataset released in 2019 by Cao et al. [38] was used in our study. The dataset contains 62 EEG data sets collected from 27 subjects (aged between 22 and 28) during 2005–2012. The participants were students or staff from the National Chiao Tung University.

A sustained-attention driving task was implemented in a simulated VR driving simulator. The participants were asked to drive and keep the car in the center of a lane. In order to simulate small changes in road curvature or stones on the way that makes car drift, lane-departure events were randomly introduced which drifted the car to the left or the right from the central lane. The participants were asked to keep their attention during the whole experiment and respond immediately to the lane-departure events by steering the car back to center of the lane. Each lane-departure event was defined as a "trial", which includes a baseline period, deviation onset (when the car starts to drift), response onset (when the participant starts to turn the wheel) and response offset (when the car returns to the central lane). The unexciting and monotonous task could easily make participants feel drowsy.

The EEG signals were sampled at a rate of 500 Hz during the whole experiment by a wired EEG cap with 30 EEG electrodes and 2 reference electrodes, which were placed according to a modified international 10–20 system. Both the raw and the processed datasets are available online.

*B. Data extraction*

The pre-processed version of the dataset available from [39] was used in this study. The dataset has already been pre-processed by its authors in the following steps:

- The raw EEG signals were filtered by 1-Hz high-pass and 50-Hz low-pass finite impulse response (FIR) filters.
- For artefact rejection, apparent eye blink contamination was manually removed. Ocular and muscular artefacts were removed by the Automatic Artifact Removal (AAR) plug-in provided in EEGLAB.

We further down-sampled the original data from 500 Hz to 128 Hz and extracted 3s-long EEG samples prior to deviation onset for each trail. The EEG data from the Oz channel were used in the study, since the channel was found to contain the most distinctive features in differentiating drowsy and alert EEG signals [40]. Each sample has 1 (channel) × 384 (sample points) dimensions.

Next, we followed the procedures described in [41] to label the EEG samples. The extent of instant drowsiness was measured by the local reaction time (RT), which is defined by the length of the interval between onset of car drift and participant's response. In addition, another index called global-RT was introduced to measure the level of drowsiness in a relatively longer period—it is calculated by averaging the RTs across all trials within a 90-second window before the onset of deviation [42]. The method takes individual performance into account by defining the baseline 'alert-RT' as the $5^{th}$ percentile of local RTs across the entire session. Trials were labeled as 'alertness' when both local and global-RT were shorter than 1.5 times alert-RT, while labeled as 'drowsiness' when both local and global RT were longer than 2.5 times alert-RT. In this way, the transitioning state with moderate performance can be excluded. The numbers of samples that can be extracted with labels from each session are listed in Appendix.

We further filtered the samples in the following procedures:

- Step 1: Sessions with less than 50 samples of either class were discarded.
- Step 2: For multiple sessions from the same subject, we selected the one with the most balanced class distribution.
- Step 3: Samples from each session were further balanced by choosing from the majority class the most representative ones with shortest (for alert class) or longest (for fatigue class) local RTs.

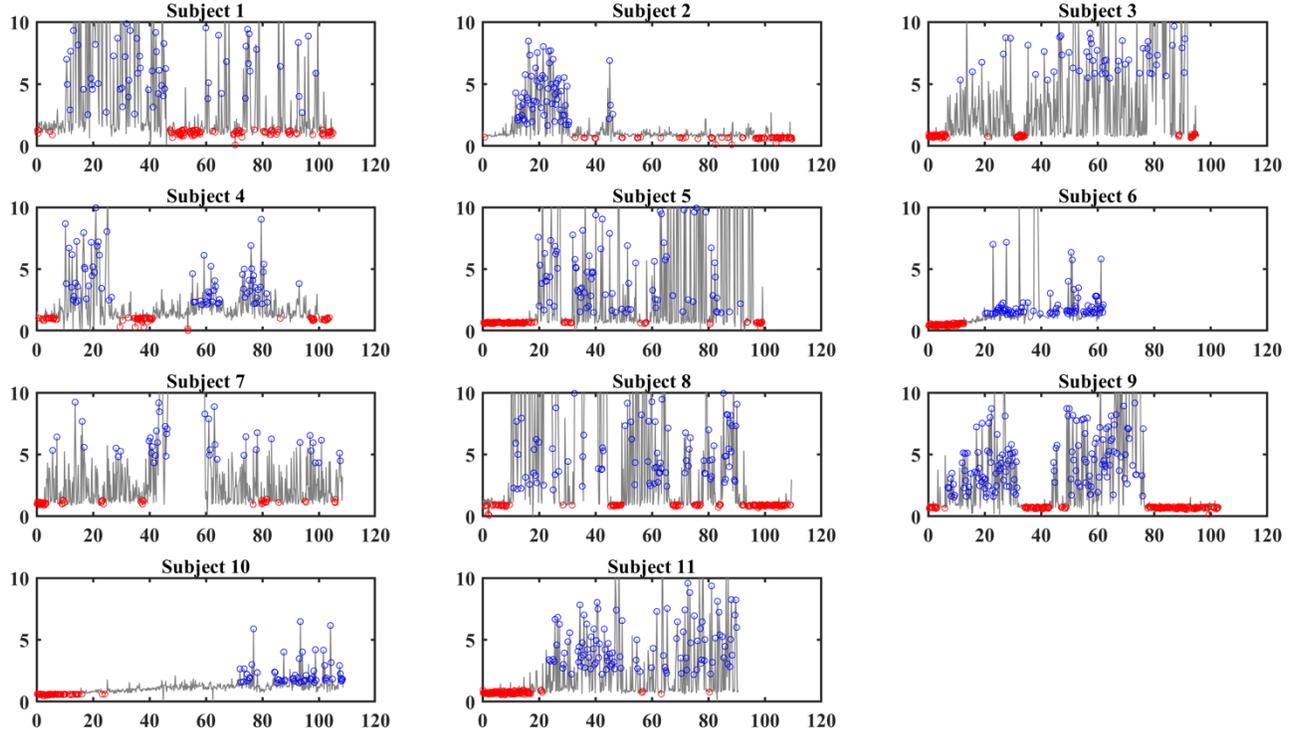

**Figure 1**. Distribution of the extracted samples against experiment time for 11 subjects. x-axis is the experimental time (minutes) and y-axis is the response time (seconds). The red circles indicate where samples in alert states are extracted, while the blue circles indicate where samples in drowsy states are extracted.

The first step excludes sessions with highly skewed class distribution. The second step makes samples from different subjects loosely balanced, so that the classifier will not favor to predict data from a specific subject. The third step ensures the classifier are trained with samples from balanced classes. The conditions above encourage the classifier to learn key features that can best discriminate between drowsiness and alert samples.

Table 1. Number of extracted samples from each eligible subject

| Subject ID | File Name | Sample Number | |
| --- | --- | --- | --- |
| | | Alert | Drowsiness |
| 1 | s01_061102n.set | 94 | 94 |
| 2 | s05_061101n.set | 66 | 66 |
| 3 | s22_090825n.set | 75 | 75 |
| 4 | s31_061103n.set | 74 | 74 |
| 5 | s35_070322n.set | 112 | 112 |
| 6 | s41_080520m.set | 83 | 83 |
| 7 | s42_070105n.set | 51 | 51 |
| 8 | s43_070205n.set | 132 | 132 |
| 9 | s44_070325n.set | 157 | 157 |
| 10 | s45_070307n.set | 54 | 54 |
| 11 | s53_090918n.set | 113 | 113 |
| | Total | 1011 | 1011 |

In this way, we finally got 2022 samples in total from 11 different subjects. The number of samples for each subject/session is shown in Table 1. Distribution of the selected samples against experiment time in each session is displayed in Figure 1. The x-axis in Figure 1 is the duration of the session and the y-axis denotes the reaction time. Dramatical fluctuations of reaction time can be observed in each session, which reflects the drowsiness when subjects alternated between sleepy and awake—they were struggling to sustain their attention in order not to fall asleep. The observation justifies the necessity to take global-RT into consideration when labeling the samples, since local quick-response may also happen in the global drowsy states. The selected samples with labels of alertness majorly fall into the period when subjects had constant quick response, while the samples with labels of drowsiness mostly fall inside period when they were not able to sustain their wakefulness.

IV. METHOD

*A. Network design*

We use CNN as our model, considering its wide applications in processing EEG signals and other forms of time-series data. A typical CNN structure consists of alternate convolutional and pooling layers with fully connected layers at the end. Some extension layers such as batch normalization [43] and dropout [44] have also been found useful to accelerate convergence of the model and prevent overfitting. Our objective is to build a CNN model for discovering and visualizing cross-subject EEG features for drowsiness detection from single-channel signals. We use a compact design of the network with only essential components for ease of interpretation. The structure of the model is described below.

- The first layer consists of 32 one-dimensional convolutional filters of size 64 with stride of 1. Applying these 32 convolutional filters on the one-dimensional EEG data results in a multivariate signal of 32 dimensions. We set the length of a filter as half the sampling rate (128 Hz) of the data so that it can capture frequency information of the range above 2Hz, which covers the essential EEG frequencies. The convolutional layer is followed by a batch normalization layer and an activation layer. The batch normalization operation [43] can remove internal covariate shift by normalizing each feature dimension across the mini-batch. Our preliminary tests show that adding the batch normalization layer after the convolutional layer accelerates convergence and improves stability of the network. We used the ELU function [45] in the activation layer to add non-linearity transformation to the data.
- Finally, we use a Global Average Pooling (GAP) layer at the end of the network followed by the Softmax function. The global pooling operation is equivalent to applying a pooling operation with the window size equal to the size of the signal outputted by the activation layer, and it reduces the filtered signals of 32 dimensions into 32 feature points. This layer dramatically reduces the parameters of the model and thus can effectively prevent over-fitting during model training.

The structure of the model is shown in Figure 2. Different from most exiting models using standard average pooling layers combined with dropout layers in the CNN structure, we use the GAP layer proposed by Lin et al. [11] instead, in order to facilitate visualization of the learned features with the Class Activation Map (CAM) method [12]. In comparison to fully connected layer as commonly used in EEG signal processing [46], the GAP layer can be interpreted as a structural regularizer which directly links the feature maps to the classification categories and significantly reduces the amount of model parameters. In addition, the layer is robust to spatial translations of the input, so that microstructures nested in EEG signals (e.g., alpha spindles [47]), which are indicators of drowsiness, can be sensitively detected. In comparison to existing deep learning models [28, 34], the novelty of the model lies in its compact design, which enables it to deal with datasets with relatively small volume of data (∼2k). The use of the GAP layer also accelerates convergence of the model and enables direct interpretation of the classification results with the CAM method.

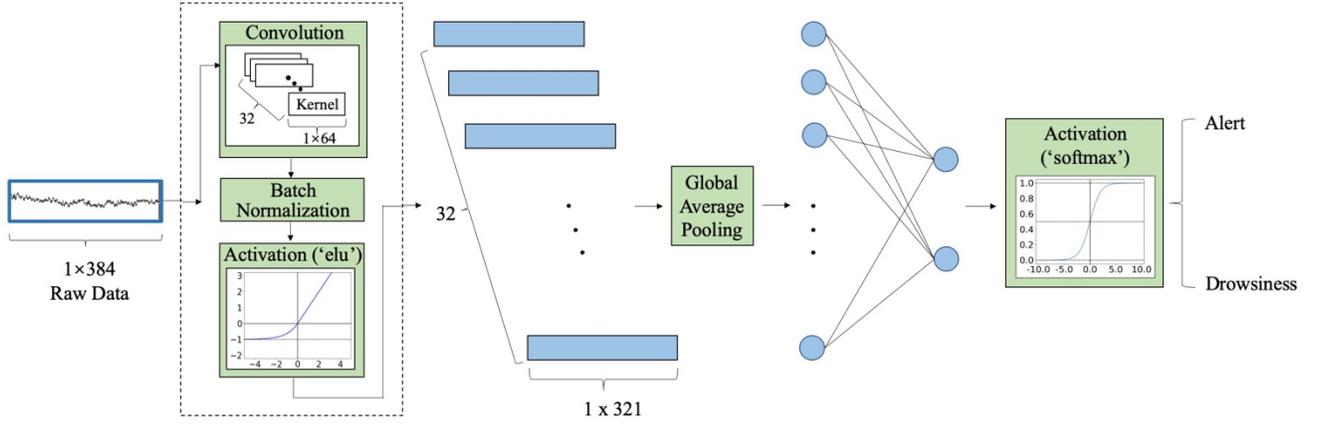

Figure 2. The structure of the proposed compact CNN model

*B. Visualization with the CAM method*

The CAM method allows us to visualize the discriminative regions of the input EEG signals found by the model as evidence to make decisions, so that we are able to discover common features across different subjects that can be signs of drowsiness. In this section, we illustrate how the CAM method is implemented on the proposed network.

For a given input signal $\mathbf{x} = \{x_i\}$ ($i = 1, 2, ..., 384$), suppose activation of nodes in the Activation ('elu') layer of model is $h^a_{k,j}$ where $k = 1, 2, ..., m$, is the feature dimension index and $j = 1, 2, ..., n$, represents the spatial index. The superscript "a" is short for "*activation*" indicating the layer name. The output dimension of this layer is 321 ($n$) x 32 ($m$), since the original signal of length 384 is reduced by 63 and sent to 32 different feature layers by the first convolutional layer. Suppose $h^g_k$ represents activation of the $k$-th node in the GAP layer. We have

$$h^g_k = \frac{1}{n}\sum_j h^a_{k,j} . \qquad (1)$$

Suppose $h^d_c$ represents the activation of node $c$ in the dense layer. For our case, $c = 0$ or 1, which corresponds to the alert and drowsy states, respectively. We have

$$h^d_c = \sum_k w_{k,c} h^g_k + b_c , \qquad (2)$$

where $w_{k,c}$ is the weight corresponding to class $c$ for node $h^g_k$ and $b_c$ is the bias for class $c$.

$h^d_c$ can be viewed as the final activation of the network to classify the input sample into class $c$. By taking Equation (1) into Equation (2), we have

$$h^d_c = \sum_k w_{k,c} h^g_k = \sum_k w_{k,c} \sum_j h^a_{k,j} = \sum_j \sum_k w_{k,c} h^a_{k,j} = \sum_j M_c(j) , \qquad (3)$$

where

$$M_c(j) = \sum_k w_{k,c} h^a_{k,j} . \qquad (4)$$

In Equation (4) $b_c$ and $\frac{1}{n}$ are ignored for simplicity. $M_c(j)$ in Equation (4) is the activation map with the length of 321 for class $c$, since sum of its entries is equivalent to $h^d_c$, which is the final activation for class $c$. In the original paper [12], the final heatmap indicating which parts of the input sample contribute most to the classification is obtained by directly up-sampling $M_c(j)$ to the same size as the input sample. For our model which contains only one convolutional layer, we consider an alternative way to align $M_c(j)$ with the contributing area of the input signal more accurately by shifting $M_c(j)$ to the central part of the input signal. Specifically, suppose the length of the convolutional kernel in the first layer is denoted as $l$ ($l = 64$), and the length of the input signal is denoted as $L$ ($L = 384$). We calculate

$$H_c(i) \begin{cases} \dfrac{2i-2}{l-2} M_c(1) & for\ 1 \leq i < \dfrac{l}{2} \\ ReLU\left(M_c\left(i - \dfrac{l}{2} + 1\right)\right) & for\ \dfrac{l}{2} \leq i \leq L - \dfrac{l}{2} \\ \dfrac{2L - 2i}{l} M_c(L - l + 1) & for\ L - \dfrac{l}{2} < i \leq L \end{cases} \quad (5)$$

The final heatmap is obtained by normalizing $H_c(i)$. In Equation (5), the negative activation of $M_c(j)$ is muted by the ReLU function, since we are only interested in the image region that can generate positive activation for the class. The extra parts of $H_c(i)$ after alignment with $M_c(j)$ are interpolated linearly between 0 and the two end values of $M_c(j)$.

V. COMPARISON DESIGN

*A. Comparison methods*

We consider comparing our model with both state-of-art deep learning methods and conventional methods.

*1) Deep CNN*

The first deep learning model for comparison is the Deep CNN model proposed by Fahimi et al. [34]. The model can differentiate between attention and non-attention states by learning their frequency components from single-channel EEG signals. The first layer of the model is a convolutional layer with 60 filters, which is followed by a max pooling layer and another two convolutional layers with 40 and 20 filters, respectively. The learned features are flattened and go through a dropout layer with probability of 0.2. It is followed by a fully-connected layer with size 100 and another dropout layer with probability of 0.3. The final layer is an activation layer with the Softmax function.

Since the original network is designed to process EEG signals with 85.33 Hz instead of 128 Hz used in the paper, we multiplied the length and stride of the first convolutional layer by 1.5 in order to fit our data. Specifically, the kernel size and stride size are changed from (1, 4) and (1, 2) to (1, 6) and (1, 3), respectively. We set all the other parameters the same as described in the original paper.

*2) EEGNet*

Another deep learning model we used for comparison is EEGNet proposed by Lawhern et al. [28]. The model was designed for a general EEG signal classification purpose and tested on several different BCI datasets under both intra-subject and cross-subject conditions. The model contains three blocks. In the first block, two steps of convolution (temporal and depthwise convolution) are performed on the data, which is inspired by the filter bank common spatial patterns (FBCSP) algorithm [48]. The first convolution step encourages the model to learn temporal features, while the model is expected to extract useful spatial features from the data in the second convolution step. The second block contains a separable convolutional layer, which can learn individual temporal feature maps. In the final classification block, features are passed directly to the dense layer, while the fully-connected layer is discarded for reducing the parameters.

Our implementation of the model is based on a slight modification of the original models accessible from [49]. We deleted the depthwise layer and the following batch normalization layer in the original implementation which were designed to learn the spatial information from the multi-channel EEG data while they are not applicable to our single-channel EEG data. The EEGNet has three hyper-parameters and they are $F_1$—number of temporal filters, $D$—number of spatial filters, and $F_2$—number of pointwise filters. In the original paper, two configurations of the network EEGNet-4,2 (with $F_1 = 4$, $D = 2$) and EEGNet-8,2 (with $F_1 = 8$, $D = 2$) were used and $F_2$ was set as the number of channels after temporal and depthwise convolution. The model we use is based on a modification of EEGNet-8,2, since the accuracy of EEGNet-4,2 was not higher than that of EEGNet-8,2 across several datasets in their experiment. We set $F_1 = 8$ and $F_2 = 8$ in our study, while parameter $D$ was not involved since we do not use the spatial layers of the network.

*3) Conventional method*

Band power features are extracted from the Oz channel using the Welch's method and relative powers of Delta (1–4 Hz), Theta (4–8 Hz), Alpha (8–12 Hz) and Beta (12–30Hz) bands are calculated. In this way, each EEG sample signal is converted to a vector with dimension of 4. Different classifiers are tested, which include Decision Tree (DT), Random Forest (RF), k-nearest neighbors (KNeighbors), Gaussian Naive Bayes (GaussianNB), Logistic Regression (LR), Linear Discriminant Analysis (LDA), Quadratic Discriminant Analysis (QDA), and SVM.

## B. Implementation details

The codes were implemented with Python 3.6.6. The experiments were conducted on the Windows 10 platform powered by an NVIDIA GeForce GTX 1080 graphics card. The proposed model was implemented with the Pytorch Library. The baseline deep learning models of Deep CNN and EEGNet were implemented with Keras API and TensorFlow 1.13.1 as the backend. Adam method [50] was used as the optimizer with learning rate of 0.01 and other default parameters ($\eta = 0.001$, $\beta_1 = 0.9$, $\beta_2 = 0.999$, $\varepsilon = $ 1e-07) were used. The conventional baseline methods were implemented with default parameters using the sklearn package [51].

## VI. RESULTS

### A. Cross-subject classification accuracy

In this comparison test, leave-one-subject-out cross-subject validation is carried out on the dataset. Specifically, the data from one subject are used for testing, while the data from all other subjects are used for training. The process is iterated until every subject has served once as the test subject. The purpose of conducting cross-subject validation is to find out whether the model can capture common features from the data pool consisting of diverse signals from different subjects with a low signal-to-noise rate, which is important for building a driver drowsiness monitoring system.

Considering the neural networks are stochastic so that the results are not consistent for each training, we repeat evaluation of each model on each test subject for 10 times. In each evaluation, we randomly initialize the network parameters and train the model from 1 to 50 epochs with a batch size of 50. In this way, we get 10 (times) x 11 (subjects) = 110 folds for each epoch, and the average accuracies are reported. The change of accuracy against training epoch is shown in Figure 3.

From Figure 3, we can see that the proposed model has an overall better performance than the other two models. It achieves the highest average accuracy of 73.22% among the three models after 6 epochs of training. The accuracy stabilizes above 71% in the first 50 epochs, which is higher than those of Deep CNN and EEGNet. The results indicate that the proposed model has better capability in extracting common EEG features related to drowsiness shared by different subjects than the other two models. Although the accuracy of EEGNet increases rapidly in first 3 epochs to 70.48%, it drops all the way afterwards and finally stabilizes at around 67.5%, which is lower than that of the proposed model. In comparison to the other two models, it takes the Deep CNN model longer time to converge to a lower accuracy than that of the proposed model.

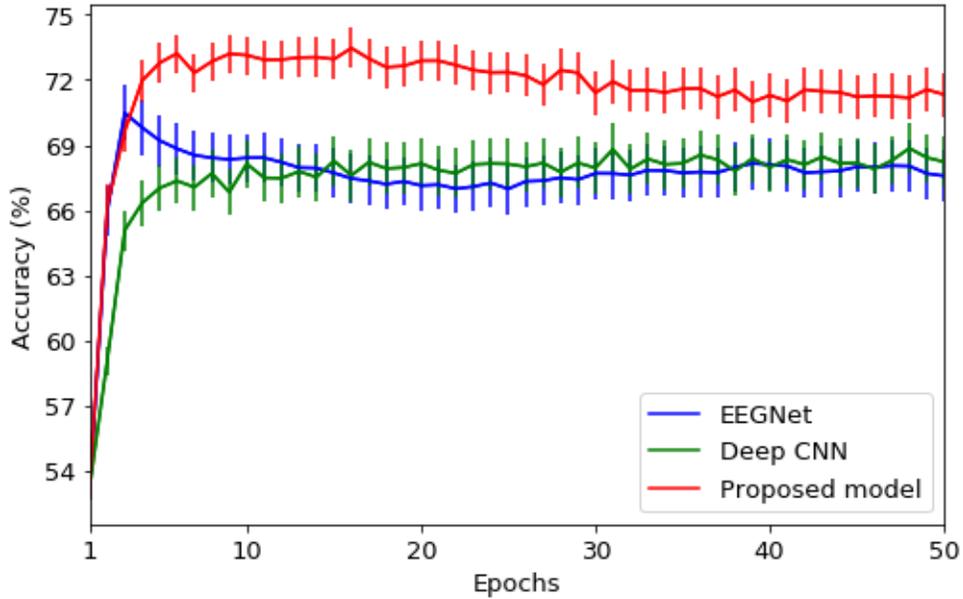

**Figure 3**. The average cross-subject classification accuracies and standard errors of Deep CNN, EEGNet and the proposed model for training epochs from 1 to 50.

Next, we compare the accuracies between the proposed method with the 8 conventional baseline methods and the results are shown in Table 2. The mean accuracy of the proposed model for each subject is obtained by averaging over 10 repetitions after 6 training epochs. Results show that the proposed model achieves the highest mean accuracy of 73.22% among the 9 methods, which is followed by 69.27% of SVM and 67.77% of LR. Paired t-tests show that the mean accuracy of the proposed model is significantly higher than DT ($p \approx 0$), RF ($p \approx 0$), KNeighbors ($p \approx 0$), GaussianNB ($p < 0.05$) and QDA ($p < 0.05$), while there is no significance on the mean accuracies between the proposed model and LR, LDA and SVM. It can also be observed that most of the highest individual accuracies (7 out of 11) are achieved by the proposed method.

**Table 2.** Comparison on the accuracies (%) between the proposed model and conventional methods. The average accuracy of the proposed model after 6 epochs is reported.

| Subject ID | DT | RF | KNeighbors | GaussianNB | LR | LDA | QDA | SVM | Proposed model |
|---|---|---|---|---|---|---|---|---|---|
| 1 | 60.11 | 70.21 | 68.09 | 77.13 | 71.28 | 75.00 | **77.66** | 77.13 | 77.45 |
| 2 | 47.73 | 48.48 | 45.45 | 41.67 | 43.94 | 43.94 | 39.39 | 46.21 | **52.80** |
| 3 | 50.67 | 46.00 | 46.00 | 52.67 | 52.67 | 49.33 | 51.33 | 49.33 | **63.47** |
| 4 | 58.78 | 50.68 | 59.46 | 57.43 | 55.41 | 50.68 | 58.11 | 61.49 | **76.22** |
| 5 | 70.09 | 62.50 | 64.29 | 62.50 | 65.18 | 64.29 | 60.27 | 68.75 | **76.52** |
| 6 | 74.10 | 80.12 | 76.51 | 81.93 | 86.75 | **87.95** | 83.13 | 85.54 | 77.11 |
| 7 | 51.96 | 57.84 | 59.80 | 66.67 | 66.67 | 65.69 | 64.71 | 63.73 | **67.35** |
| 8 | 63.64 | 67.80 | 69.70 | 73.86 | **79.17** | 77.65 | 71.97 | 73.48 | 71.93 |
| 9 | 69.11 | 69.75 | 73.57 | 75.48 | 73.25 | 76.11 | 78.34 | 81.21 | **88.25** |
| 10 | 65.74 | 72.22 | 75.00 | 87.04 | 86.11 | **87.96** | **87.96** | 86.11 | 81.67 |
| 11 | 55.75 | 57.08 | 59.73 | 65.49 | 65.04 | 65.04 | 65.04 | 69.03 | **72.65** |
| Average | 60.70 | 62.06 | 63.42 | 67.44 | 67.77 | 67.60 | 67.08 | 69.27 | **73.22** |

In order to understand how each part of the model influences its performance and whether the current model is optimal in design, we test performance of the model when a single component of the model is modified or deleted. Specifically, we test two variations of the model without the activation layer (2$^{nd}$ layer) and the batch normalization layer (3$^{rd}$ layer). In addition, we explore the benefits of using the GAP layer (4$^{th}$ layer) by replacing it with standard average pooling layers having pooling sizes of 40 and 80, respectively. In order to keep the same number of nodes as that of the original model, we add an additional dropout layer to these two variations with dropout rate (probability of the node to be dropped out) of 0.875 and 0.75, respectively. The comparison results are shown in Figure 4.

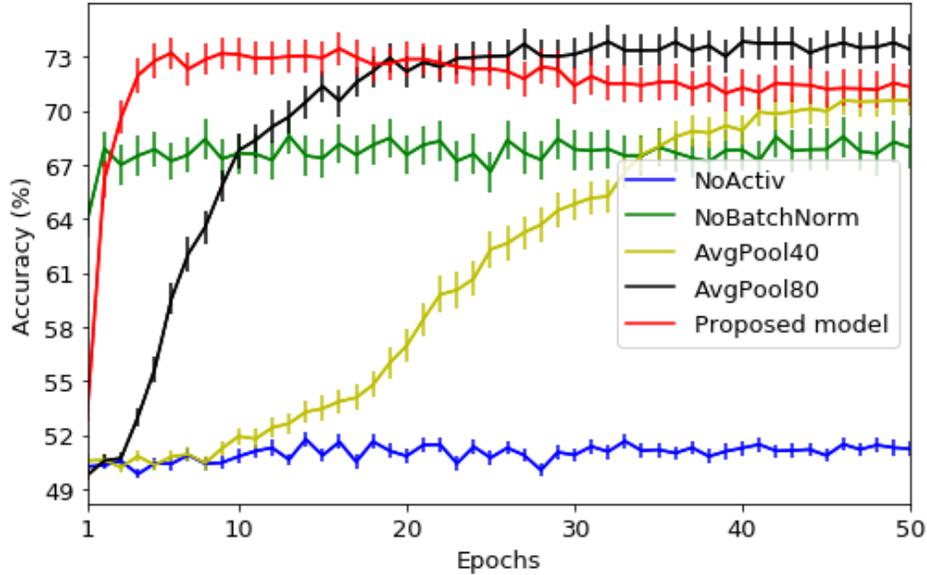

**Figure 4.** The average cross-subject classification accuracies and standard errors of the proposed model and its variations against training epochs from 1 to 50. In the figure, "NoActiv" and "NoBatchNorm" represent the models without the activation and the batch normalization layers in comparison to the original model, respectively. "AvgPool40" represents the model when the GAP layer of the original model is replaced with an average pooling layer with pooling size of 40 and a dropout layer after it with a dropout rate of 0.875. "AvgPool80" represents the model when the GAP layer of the original model is replaced with an average pooling layer with pooling size of 80 and a dropout layer after it with a dropout rate of 0.75.

From Figure 4, we can see that the ELU activation layer plays an important role in the model. The average accuracy drops more than 20% when the ELU activation layer is removed from the original model. Batch normalization is also an important component of the network since it accounts for around 6% to the average classification accuracy. In fact, the batch normalization layer was proposed to solve the internal covariate shift problem by normalizing the activations from the previous layer with zero mean and unit variance. For our model, the layer serves the function of removing the individual-level feature drift converted from the input signals of the test subject. By comparing the GAP layer with standard average pooling layers, we can find that the GAP layer allows a faster convergence of the model. The original model reaches the peak mean accuracy of 73.22% after 6 training epochs, while the model AvgPool80 reaches its peak mean accuracy of 73.85% after 40 epochs. It takes even longer training epochs for the model AvgPool40 to converge. We compared the highest mean accuracies for the 11 subjects of the original model and the AvgPool80 model with paired t-test ($df$ =10), and the result shows that there is no significant difference between them.

*B. Visualization on the learned patterns*

Having obtained the previous results, we consider exploring shared EEG features across different subjects that have been learned by the model as indicators of alert and drowsy states. We analyzed the results and show some representative samples from different subjects in Figure 5 and Figure 6 to interpret the common patterns learned by the model.

For samples classified with a high likelihood of drowsy labels, we have observed two common patterns with regard to their frequency components. Samples from the first category commonly have a high portion of Alpha waves, e.g., Figure 5(a) and Figure 5(b), while samples from the second category commonly have a high portion of slow waves in the Delta and Theta frequency bands, e.g., Figure 5(c) and Figure 5(d). For the first sample shown in Figure 5(a), it is revealed by the CAM method that the model has captured a spindle-like micro-structure in the first half part of the sample as evidence for drowsiness. By observation on its corresponding multi-channel signals, we can see the spindle is generated from the parietal-occipital areas of the cortex and propagates as far as to the central area of the cortex. For the second sample shown in Figure 5(b), a similar spindle has been identified but with stronger amplitude and higher relative power in the Alpha band. It can also be observed that the spindle is generated from the similar region of the cortex as the previous one, while it propagates to a wider area of the cortex with a longer duration. Actually, these captured microstructures, which can be characterized with an arrow frequency peak within the alpha band and a low-frequency modulation envelope resulting in the typical 'waxing and waning' of the alpha rhythm [52], are called "alpha spindles" and used as indictors to detect driver fatigue [47].

As it can be seen in Figure 5(c) and 5(d), the second pattern is associated with a high portion of Theta-Delta frequency activities while a low portion of activities in the Alpha and Beta frequencies. For the sample shown in Figure 5(c), it can be seen that the model has captured the rhythmic bursts in the Theta-Delta bands as evidence of the drowsy state. For the sample shown in Figure 5(d), the similar type of bursts has been recognized but from two separable parts of the sample signal. In fact, these bursts in the Theta–Delta frequency band, or called "drowsy bursts", have been found to frequently appear in EEG signals during drowsiness [53].

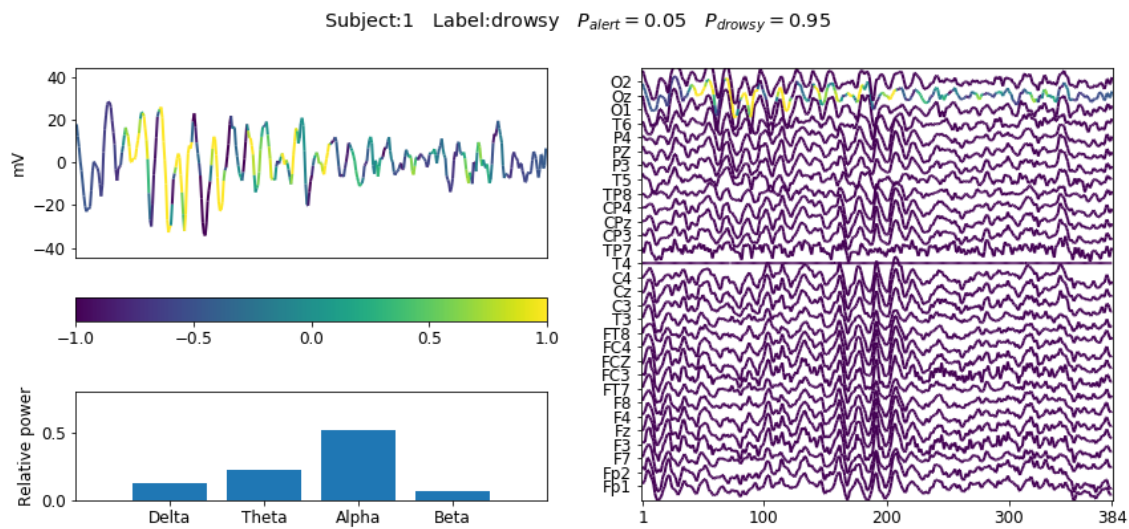

(a)

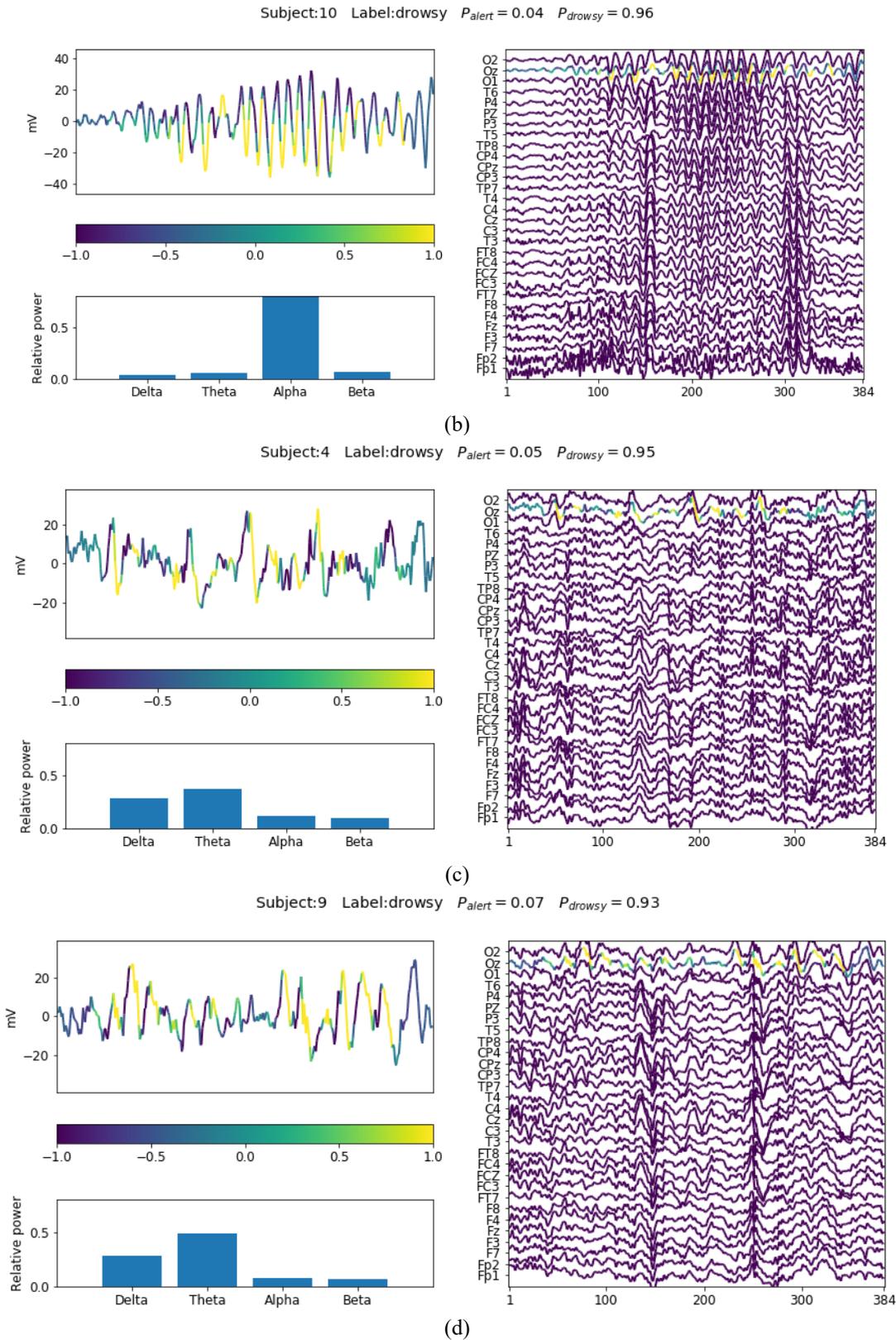

**Figure 5**. Visualization of learned patterns on selected samples that are correctly classified by the network with high likelihoods of the drowsy label. The subject ID, sample label, likelihood output by the model for alert and drowsy labels are shown on top of each sub-figures. The sample signal from Oz channel is shown in the left top part of each sub-figure. The contributing regions to classification are indicated by the line color converted from the CAM method. The relative power on frequency bands of Delta, Theta, Alpha and Beta are calculated and shown in the left bottom part of each sub-figure. The panorama of the sample with EEG signals from all the channels is shown in the right part of each figure.

The samples classified with a high likelihood of the alert label commonly contain a high portion of brain activities in the Beta frequency band, e.g., Figure 6(a) and 6(b), or the Delta band, e.g., Figure 6(c) and 6(d). For the first category represented by samples in Figure 6(a) and 6(b), we can observe that the model has identified several short episodes from the signal that contain high-frequency waves. By observation on the corresponding multi-channel EEG signals of the samples, we are able to know that the detected high frequency waves are majorly caused by electromyography (EMG) activities due to tension of scalp muscles, which usually dominates the wakeful EEG signals [53].

For the samples shown in Figure 6(c) and 6(d), the model has identified the regions with large-amplitude drifts as alertness-related features. However, these drifts are not typical Delta waves, which dominate during deep sleep stages [54]. They are more likely caused by sensor drifts or displacements associated with subject movements happening in their wakeful states, since the drifts occur at all the channels simultaneously.

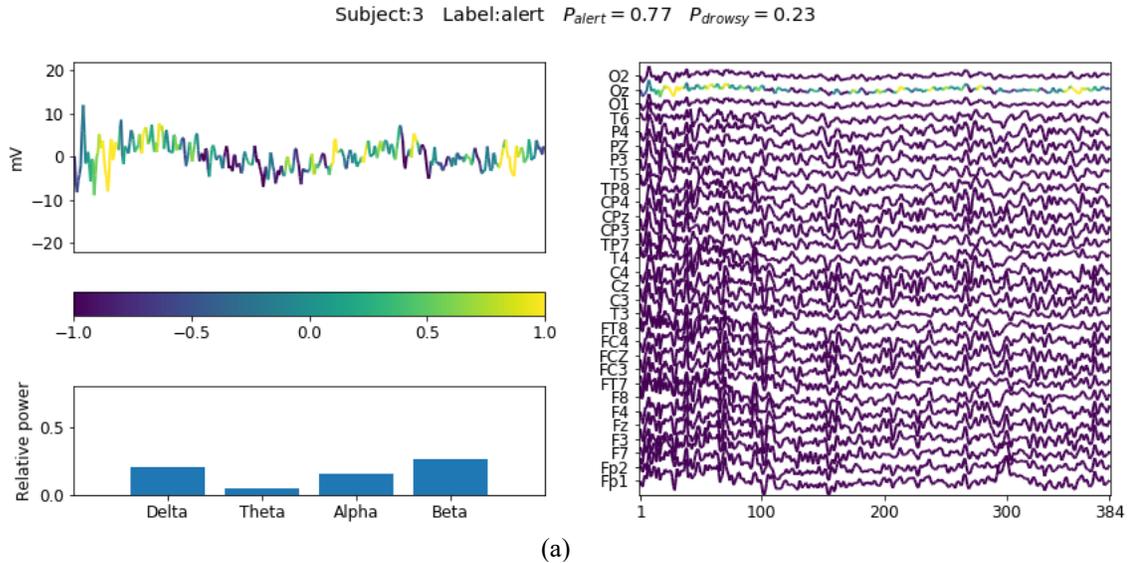

(a)

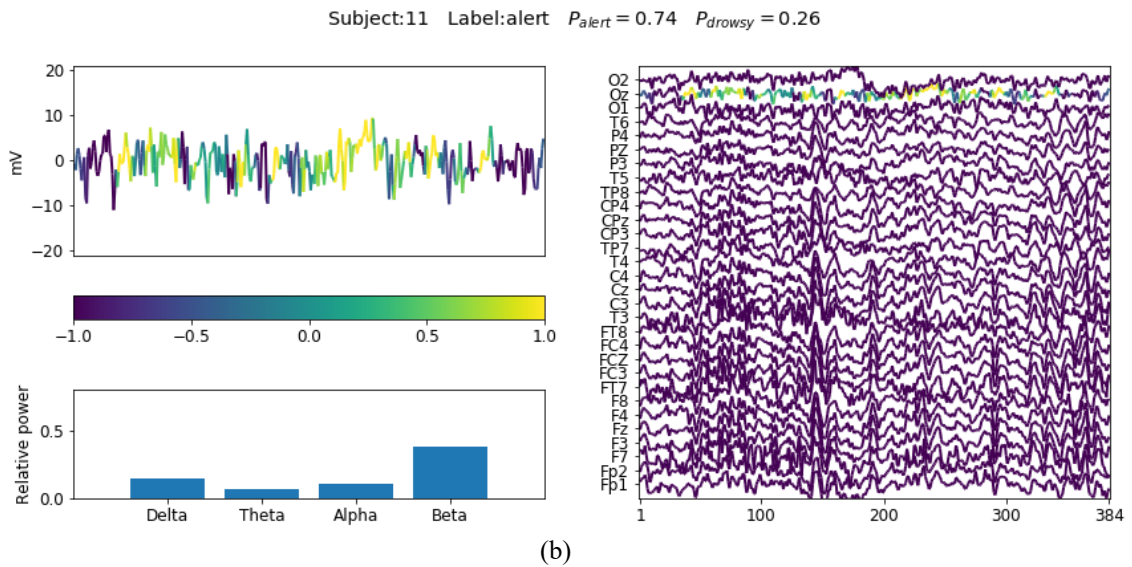

(b)

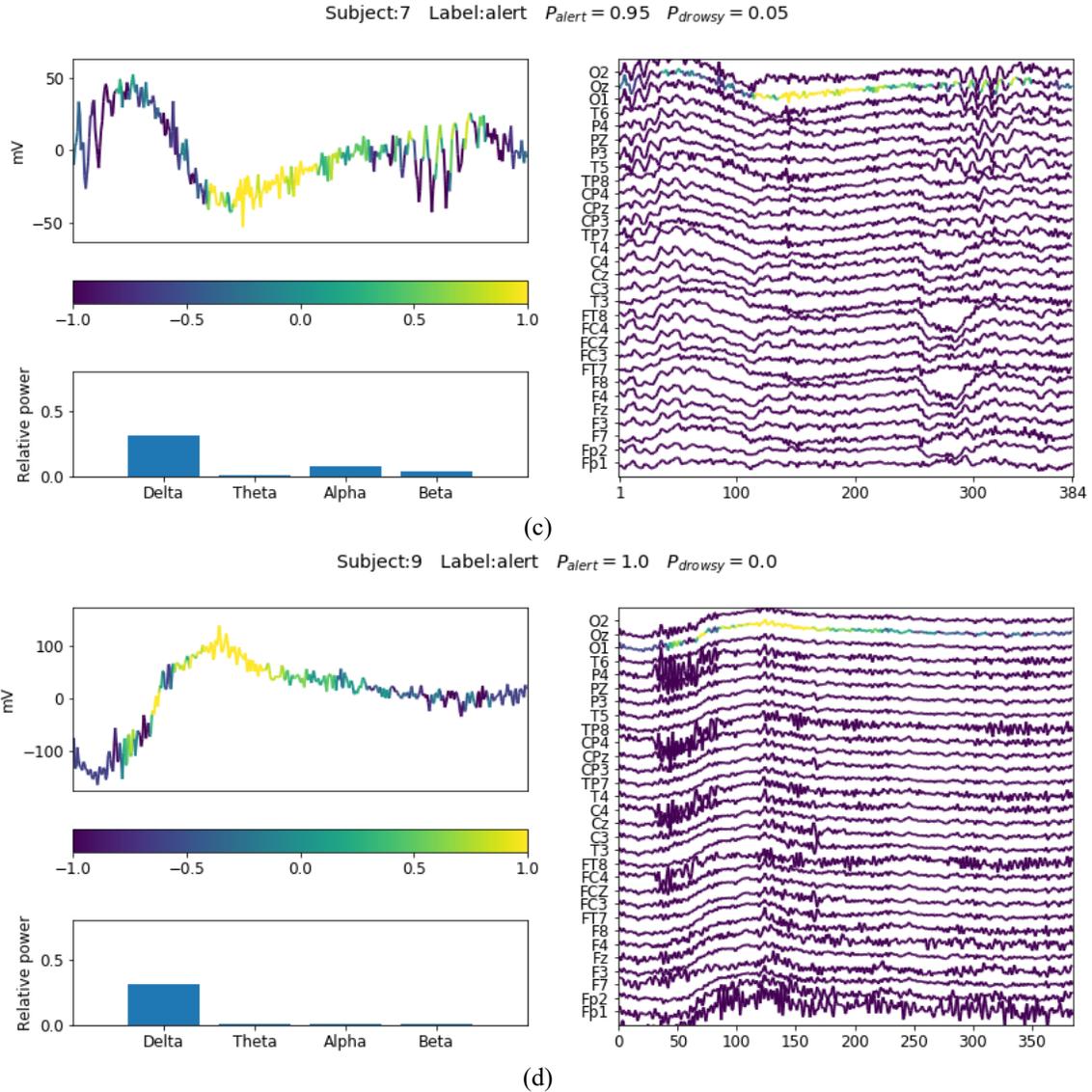

**Figure 6**. Visualization of learned patterns on selected alert samples that are correctly classified by the network with high likelihoods.

Finally, we consider investigating some wrongly classified samples with the visualization technique. From Table 2, we can see that both conventional methods and the proposed CNN model have a low classification accuracy on Subject 2. Therefore, we explore some wrongly classified samples of the subject in order to find out the reasons. The visualization results are shown in Figure 7. The first two samples shown in Figure 7(a) and 7(b) are labeled with drowsiness but are wrongly classified as alert states, while the last two samples shown in Figure 7(c) and 7(d) are labeled with alertness but are wrongly classified as drowsy states.

Similar to the samples shown in Figure 6(a) and 6(b), the first sample shown in Figure 7(a) contains a high portion of Beta waves, which is identified by the model as evidence of the alert state. As discussed previously, the Beta waves caused by EMG activities are usually present in the EEG signals when the subject is wakeful, while they are not typical features of drowsy EEG signals, as it can be seen in the samples from Figure 5. On the other hand, the first sample in Figure 7(a) does not show obvious drowsiness-related features, such as Alpha spindles or bursts in the Theta-Delta band, as observed before.

The second sample shown in Figure 7(b) is different from the first sample in the aspect that it contains not only a high portion of Beta waves but also a high portion of Alpha waves. We can see from the multi-channel signal of the sample that there indeed exists Alpha spindles in the episode generated from the frontal-central regions of the cortex, while the spindles are weakened when they have reached the occipital channels. The high portion of Alpha waves could be the reason that makes the sample to have a higher likelihood of the drowsy label in comparison to the previous sample, while the final classification is still influenced by the high portion of Beta waves contained in the sample.

For the last two samples shown in Figure 7(c) and (d), it is obvious that the model has identified Alpha spindles as strong indicators of drowsiness for classification, which is similar to the cases shown in Figure 5(a) and 5(b). The visualization results justify the classification by the model on these two samples.

Based on the observations above, we infer the low classification accuracy of samples from Subject 2 is partially attributed by the high portion of Beta waves and lack of obvious drowsiness-related features in samples with drowsy labels, which is biased from the group statistics learned by the model. On the other hand, it is also reasonable to suspect the fidelity of the alert labels for some samples that have strong drowsiness-related features revealed by the model, e.g., samples in Figure 7(c) and 7(d). It could happen that the subject was in the early drowsiness stage or was already drowsy but coincidentally responded timely when these samples were captured.

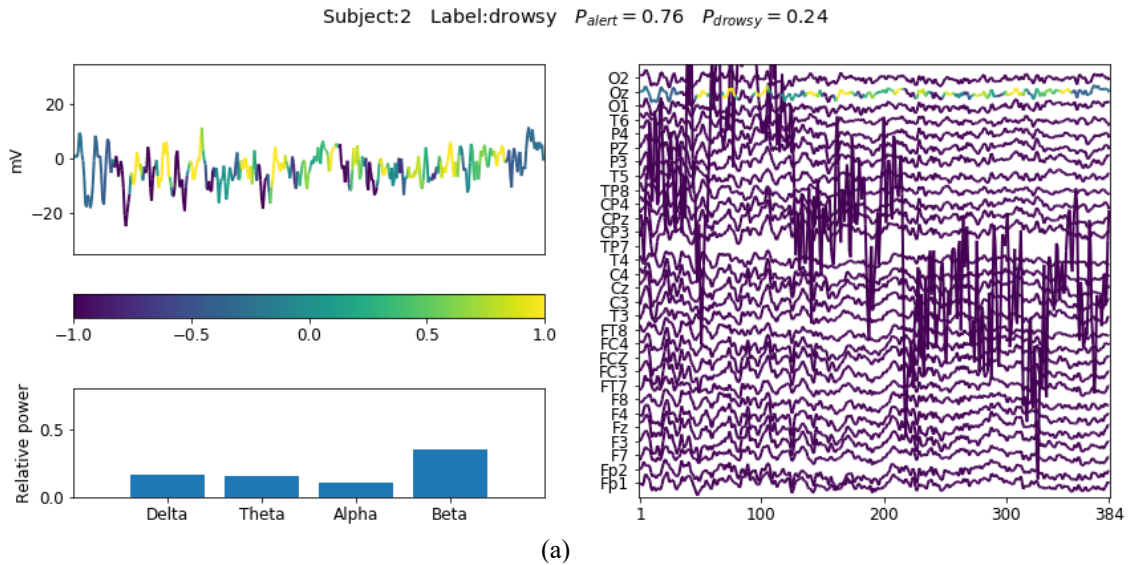

(a)

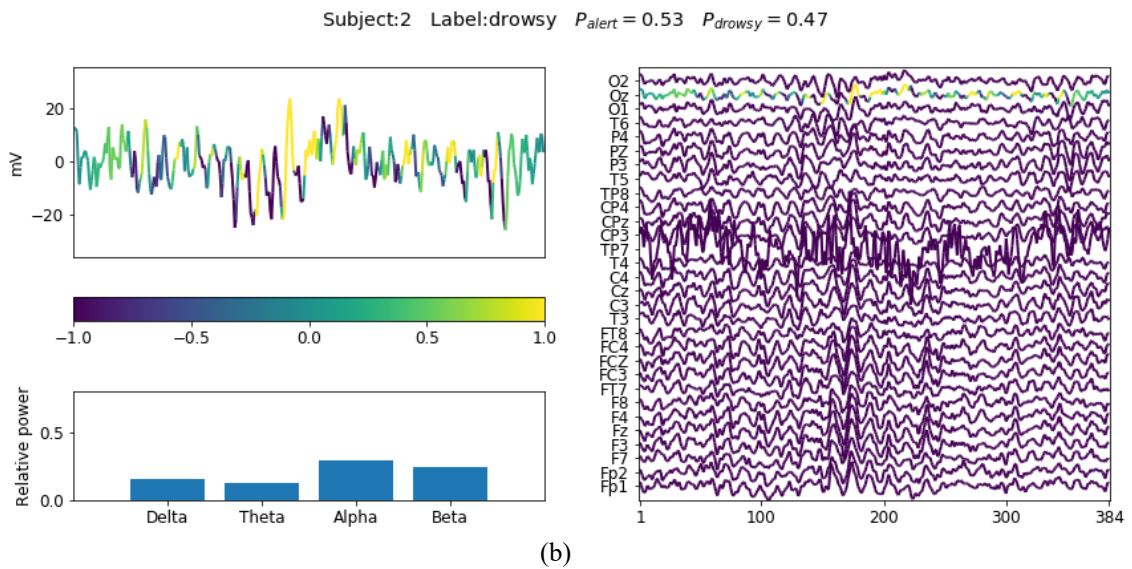

(b)

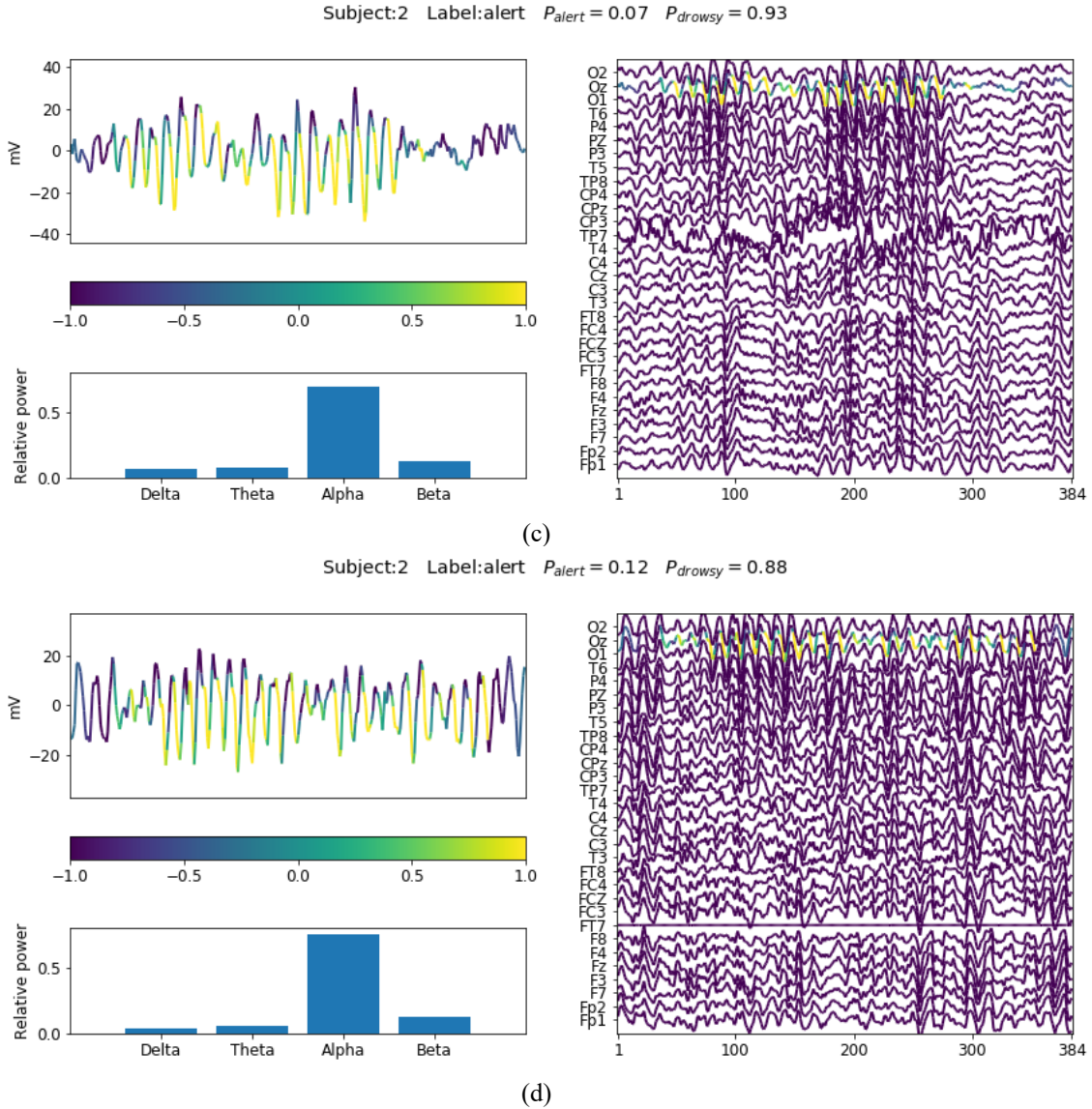

**Figure 7**. Visualization of learned patterns on selected wrongly classified samples from Subject 2.

## VII. Discussion and future works

Different from existing works that mostly treat deep learning models as black-box classifiers to recognize mental states from EEG signals, in this paper we explored using CNN model as a powerful tool to discover and visualize cross-subject EEG features that can differentiate single-channel EEG signals between alert and drowsy states. Specifically, we designed a compact model with a GAP layer in the structure allowing the importance of sample local regions for classification to be revealed with the CAM method. In fact, the visualization technique can also be extensively applied to other types of CNN models using the Grad-CAM method proposed by Selvaraju et al. [55], which can generate the class activation map for a broader range of CNN models while not only limited to the ones having a GAP layer. Specially, we evaluated the influence of the GAP layer to model performance and found the GAP layer can provide additional benefits by allowing faster convergence of the testing accuracies for the model.

The CAM method can reveal which parts of the input sample contain features that the model uses as evidence for classification, which allows us to discover the learned patterns associated with different mental states. Specifically, we have found that the model has learned biologically explainable features, e.g., Alpha spindles and Theta burst, as evidence for the drowsy state. It is also interesting to see that the model uses artifacts that are present in the wakeful EEG, e.g., muscle artifacts and sensor drifts, to recognize the alert state.

We explored potential applications of the visualization technique and considered using it for analyzing the samples wrongly classified by the network. It is interesting to find that the model can justify its classification results for some samples with the visualization results, which causes us to suspect the fidelity of their ground-truth labels. Indeed, the performance or behaviors of subjects, e.g., reaction time, may not faithfully reflect mental states of subjects in certain circumstances. It could be an interesting topic of incorporating the "network explanation" into the labeling process, instead of merely using thresholds hard-coded on behavior/performance metrics of the subjects.

The proposed model currently can only be used for offline analysis. It is not yet ready for online applications while maintaining the obtained accuracy. This is because in the evaluation phase of the current test, we send all the samples from the test subject as a bundle into the network, in order for batch normalization layer to remove individual-level feature drift of the test subject. As it can be seen in Figure 4, there is a drop of around 6% accuracy when the batch normalization layer is removed from the model, which could be caused by individual-level feature drifts. We will consider exploring other ways to calibrate the data of the test subjects, e.g., recording of a short period of EEG data in idle state before the formal session, instead of using the batch normalization layer for the purpose in our future works. In addition, the current model has only been tested on the Oz channel on a single dataset with 11 subjects as an initial attempt. We will consider testing the model on different EEG channels and datasets with more subjects in the future.

## VIII. Conclusion

In this paper, we designed a compact CNN model to recognize drowsiness from single-channel EEG data. It incorporates the GAP layer in the structure, which allows local regions of the signal that contribute most to classification to be visualized. We conducted a cross-subject test for the model on a public dataset. Results show that the proposed model not only outperforms both conventional and deep learning baseline methods but also can localize neuro-physiologically interpretable micro-structures, e.g., Alpha spindles, from EEG signals and use them to distinguish between EEG signals recorded in alert and drowsy states. The proposed model illustrates a potential direction to use CNN models as a powerful tool to discover interesting neurophysiological patterns from EEG signals.

## Acknowledgement

This research is supported by the National Research Foundation, Singapore under its International Research Centres in Singapore Funding Initiative. Any opinions, findings and conclusions or recommendations expressed in this material are those of the author(s) and do not reflect the views of National Research Foundation, Singapore.

APPENDIX

Table A1. The number of samples that can be extracted from each session of the raw dataset with the criteria described in Section III.B. The data from sessions with indices of 5, 12, 25, 30, 32, 36, 41, 43, 48, 49, 59 were used in this paper.

| Index | Filename | Subject ID | Sample number Alert | Sample number Drowsy | Index | Filename | Subject ID | Sample number Alert | Sample number Drowsy |
|---|---|---|---|---|---|---|---|---|---|
| 1 | 's01_051017m.set' | 1 | 1 | 155 | 32 | 's35_070322n.set' | 14 | 161 | 112 |
| 2 | 's01_060227n.set' | 1 | 1 | 247 | 33 | 's40_070124n.set' | 15 | 171 | 42 |
| 3 | 's01_060926_1n.set' | 1 | 12 | 41 | 34 | 's40_070131m.set' | 15 | 10 | 125 |
| 4 | 's01_060926_2n.set' | 1 | 50 | 20 | 35 | 's41_061225n.set' | 16 | 295 | 25 |
| 5 | 's01_061102n.set' | 1 | 94 | 96 | 36 | 's41_080520m.set' | 16 | 83 | 116 |
| 6 | 's02_050921m.set' | 2 | 0 | 309 | 37 | 's41_080530n.set' | 16 | 45 | 162 |
| 7 | 's02_051115m.set' | 2 | 0 | 316 | 38 | 's41_090813m.set' | 16 | 52 | 208 |
| 8 | 's04_051130m.set' | 3 | 0 | 304 | 39 | 's41_091104n.set' | 16 | 60 | 99 |
| 9 | 's05_051120m.set' | 4 | 0 | 304 | 40 | 's42_061229n.set' | 17 | 62 | 25 |
| 10 | 's05_060308n.set' | 4 | 37 | 38 | 41 | 's42_070105n.set' | 17 | 51 | 103 |
| 11 | 's05_061019m.set' | 4 | 0 | 641 | 42 | 's43_070202m.set' | 18 | 236 | 70 |
| 12 | 's05_061101n.set' | 4 | 363 | 66 | 43 | 's43_070205n.set' | 18 | 238 | 132 |
| 13 | 's06_051119m.set' | 5 | 0 | 336 | 44 | 's43_070208n.set' | 18 | 267 | 51 |
| 14 | 's09_060313n.set' | 6 | 2 | 222 | 45 | 's44_070126m.set' | 19 | 248 | 72 |
| 15 | 's09_060317n.set' | 6 | 0 | 171 | 46 | 's44_070205n.set' | 19 | 270 | 103 |
| 16 | 's09_060720_1n.set' | 6 | 0 | 237 | 47 | 's44_070209m.set' | 19 | 244 | 115 |
| 17 | 's11_060920_1n.set' | 7 | 230 | 32 | 48 | 's44_070325n.set' | 19 | 243 | 157 |
| 18 | 's12_060710_1m.set' | 8 | 0 | 324 | 49 | 's45_070307n.set' | 20 | 192 | 54 |
| 19 | 's12_060710_2m.set' | 8 | 1 | 240 | 50 | 's45_070321n.set' | 20 | 228 | 19 |
| 20 | 's13_060213m.set' | 9 | 0 | 428 | 51 | 's48_080501n.set' | 21 | 217 | 31 |
| 21 | 's13_060217m.set' | 9 | 0 | 230 | 52 | 's49_080522n.set' | 22 | 251 | 0 |
| 22 | 's14_060319m.set' | 10 | 290 | 0 | 53 | 's49_080527n.set' | 22 | 123 | 29 |
| 23 | 's14_060319n.set' | 10 | 16 | 1 | 54 | 's49_080602m.set' | 22 | 76 | 37 |
| 24 | 's22_080513m.set' | 11 | 23 | 99 | 55 | 's50_080725n.set' | 23 | 182 | 32 |
| 25 | 's22_090825n.set' | 11 | 75 | 180 | 56 | 's50_080731m.set' | 23 | 193 | 14 |
| 26 | 's22_090922m.set' | 11 | 24 | 135 | 57 | 's52_081017n.set' | 24 | 54 | 44 |
| 27 | 's22_091006m.set' | 11 | 53 | 160 | 58 | 's53_081018n.set' | 25 | 87 | 44 |
| 28 | 's23_060711_1m.set' | 12 | 88 | 14 | 59 | 's53_090918n.set' | 25 | 113 | 131 |
| 29 | 's31_061020m.set' | 13 | 113 | 10 | 60 | 's53_090925m.set' | 25 | 75 | 72 |
| 30 | 's31_061103n.set' | 13 | 118 | 74 | 61 | 's54_081226m.set' | 26 | 36 | 66 |
| 31 | 's35_070115m.set' | 14 | 85 | 101 | 62 | 's55_090930n.set' | 27 | 618 | 0 |